\documentclass[conference]{IEEEtran}
\usepackage{float}

\IEEEoverridecommandlockouts
\usepackage{amsmath,amssymb,amsfonts}
\usepackage{algorithmic}
\usepackage{authblk}
\usepackage{graphicx}
\usepackage{balance}
\usepackage[
backend=biber,
sorting=none,
url=false,
]{biblatex}
\usepackage{textcomp}
\usepackage{xcolor}
\AtNextBibliography{\footnotesize}
\def\BibTeX{{\rm B\kern-.05em{\sc i\kern-.025em b}\kern-.08em
    T\kern-.1667em\lower.7ex\hbox{E}\kern-.125emX}}
\begin{document}

\title{MM‑Open5x: Practical Conformal Electronics via Multi‑Material Multi‑Axis Additive Manufacturing}
\author{
 Iván Revenga Riesco$^{1}$}
  \author{
 Freddie Hong$^{2}$}
 \author{
 Borut Lampret$^{1}$}
 \author{
 Connor Myant$^{1}$}
 \author{
 David Boyle$^{1}$}

 \affil{Imperial College London$^{1}$, Kingston University London$^{2}$}

\maketitle

\begin{abstract}
Multi‑axis, multi‑material additive manufacturing enables the production of conformal electronic systems that are difficult to realise using planar processes and can offer enhanced performance; however, accessibility and reproducibility remain challenging. This paper presents MM‑Open5x, an open and low-cost multi‑material, multi‑axis additive manufacturing workflow for practical conformal electronics. The workflow spans design, simulation, toolpath generation, fabrication, testing, and assembly, and is released alongside complete design files, toolpaths, and representative G‑code. Representative case studies demonstrate both the benefits and limitations of multi‑axis deposition in practice, explicitly documenting common failure modes, calibration sensitivities, assembly bottlenecks, and material variability. Rather than prescribing an optimal solution, MM‑Open5x is presented as a transparent, reproducible foundation upon which a broader community can iterate, extend, and apply conformal electronics workflows across diverse application domains.
\end{abstract}


\section{Introduction}

Multi‑axis, multi‑material additive manufacturing (AM) has demonstrated the potential to enable conformal electronic systems that are difficult, or in some cases impossible, to realise using conventional planar fabrication techniques. Prior work has shown a wide range of passive and active devices leveraging AM, ranging from tunable waveguide structures to flexible pressure sensors for wearable applications \cite{shin_polymer-based_2019, bin_lee_flexible_2026}. To date, commercial adoption has been driven predominantly by Radio Frequency (RF) and microwave components for specialised use cases, e.g.,~\cite{noauthor_swissto12_2024}. Despite this progress, the broader adoption of additively manufactured electronics remains limited by challenges related to accessibility, cost-effectiveness, reproducibility and end‑to‑end practicality. 

The conventional electronics industry is inherently planar, and additive manufacturing is therefore unlikely to compete directly with this mature ecosystem in terms of scalability, cost efficiency, or manufacturing readiness. Instead, the value of AM lies in exploiting three‑dimensional fabrication to realise electronically functional structures that are geometrically complex, highly integrated, and inaccessible to planar manufacturing. 
More recently, low‑cost open‑source ecosystems such as Open5x~\cite{hong_open5x_2022} have significantly reduced the barrier to entry. However, despite a growing body of work, there remains a lack of an accessible and reproducible end‑to‑end workflow that integrates design, fabrication, and evaluation of conformal electronic systems in practice.

Responding to these challenges, this paper introduces \textbf{MM‑Open5x}, an open and low‑cost workflow to produce conformal electronics based on multi‑material, multi‑axis additive manufacturing. The key contributions of this work are as follows: i) the consolidation of design, simulation, toolpath generation, fabrication, testing, and assembly into a transparent, reproducible end‑to‑end workflow; ii) the public release of complete design artefacts, simulation models, toolpaths, and representative G‑code to enable replication and extension by other researchers;\footnote{https://github.com/BigKoala33/MM-Open5x} and iii) explicit description of practical constraints, observed failure modes and mitigation approaches, including calibration sensitivity, component assembly bottlenecks, and material variability, together with their implications for reproducibility and positioning these issues among open challenges for the community.

\section{Related work}
\begin{figure}[t]  \centering  \includegraphics[width=\linewidth]{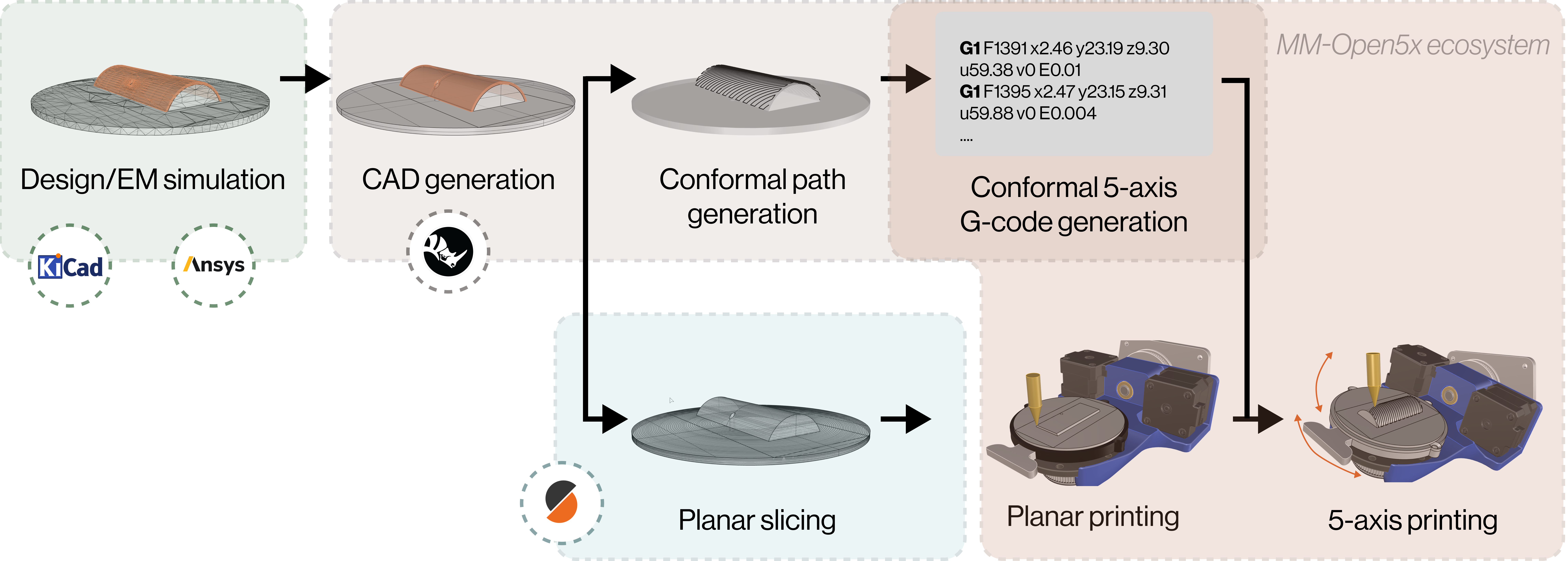}  \caption{MM-Open5x: from design to manufacture of conformal electronics.}  \label{fig:overview}\end{figure}
Open5x originated as an open‑source project aimed at democratising access to 5‑axis additive manufacturing on Material Extrusion (MEX) platforms \cite{hong_open5x_2022}. It comprises lightweight hardware modifications to conventional 3D printers together with a Rhino Grasshopper‑based 5‑axis G‑code generator that produces toolpaths normal to a target surface, enabling controlled conformal deposition. Early work demonstrated the advantages of depositing conductive filaments along surfaces rather than planarly \cite{hong_5-axis_2023}, mitigating anisotropic conductivity effects inherent to conventional MEX caused by inter‑layer interfaces at both DC and RF frequencies \cite{riesco_5-axis_2025}. Open5x has since enabled a growing body of academic work in multi‑axis additive manufacturing and conformal electronics \cite{yu_rapidly-constructed_2024, terranova_versatile_2024, ahlers_calibration_2025}. However, a notable learning curve and the lack of accessible, reproducible workflows for producing multi‑material conformal circuitry remain significant barriers to broader adoption.

\section{MM-Open5x fabrication workflow}

This section describes the workflow for fabricating conformal electronics leveraging Open5x, considering both hardware and software aspects (Fig.~\ref{fig:overview}). The workflow is intended to support low‑volume research or \textit{ad hoc} prototyping of conformal electronic structures rather than high‑throughput or production‑grade manufacturing.

\subsection{Machine Setup}

This section details the Open5x-based multi-material hardware platform used in our previous works \cite{hong_5-axis_2023, riesco_5-axis_2025, revenga_riesco_multi-axis_2025}. Some additional modifications are introduced to improve the fabrication of conformal electronics from the original designs.

\subsubsection{Tool-changer}
The tool-changer is configured with two tool heads, both based on Bowden-driven E3D Hemera extruders. The first extruder is used for the substrate material with a 0.4~mm nozzle, while the second is loaded with Electrifi using a 0.25~mm nozzle. Cooling is limited to the extruder (pre-heatbreak), with no active cooling applied to the printed part. While this constrains maximum print speed, the absence of cooling hardware near the nozzle enables printing on highly convex surfaces and reduces collision risk. The system does not include a nozzle wiper or dedicated calibration probe.

\subsubsection{MM-Open5x Gantry}
The gantry modifications are minimal, consisting primarily of an increased print area and the integration of a calibration element derived from the Voron 0 Open5x implementation. A full 5-axis calibration procedure is employed (details and G-code are provided in the repository), enabling identification of the true axis of rotation of the V-axis (the axis to spin the bed). The calibration process does not account for U–V axis misalignment or axial tilts introduced during manufacturing or assembly, which can affect print accuracy at large rotations. 
Accurately characterising these effects remains a significant challenge.

\subsubsection{Nichrome-heated nozzle extender}

\begin{figure}[t]
    \centering
    \includegraphics[width=0.5\linewidth]{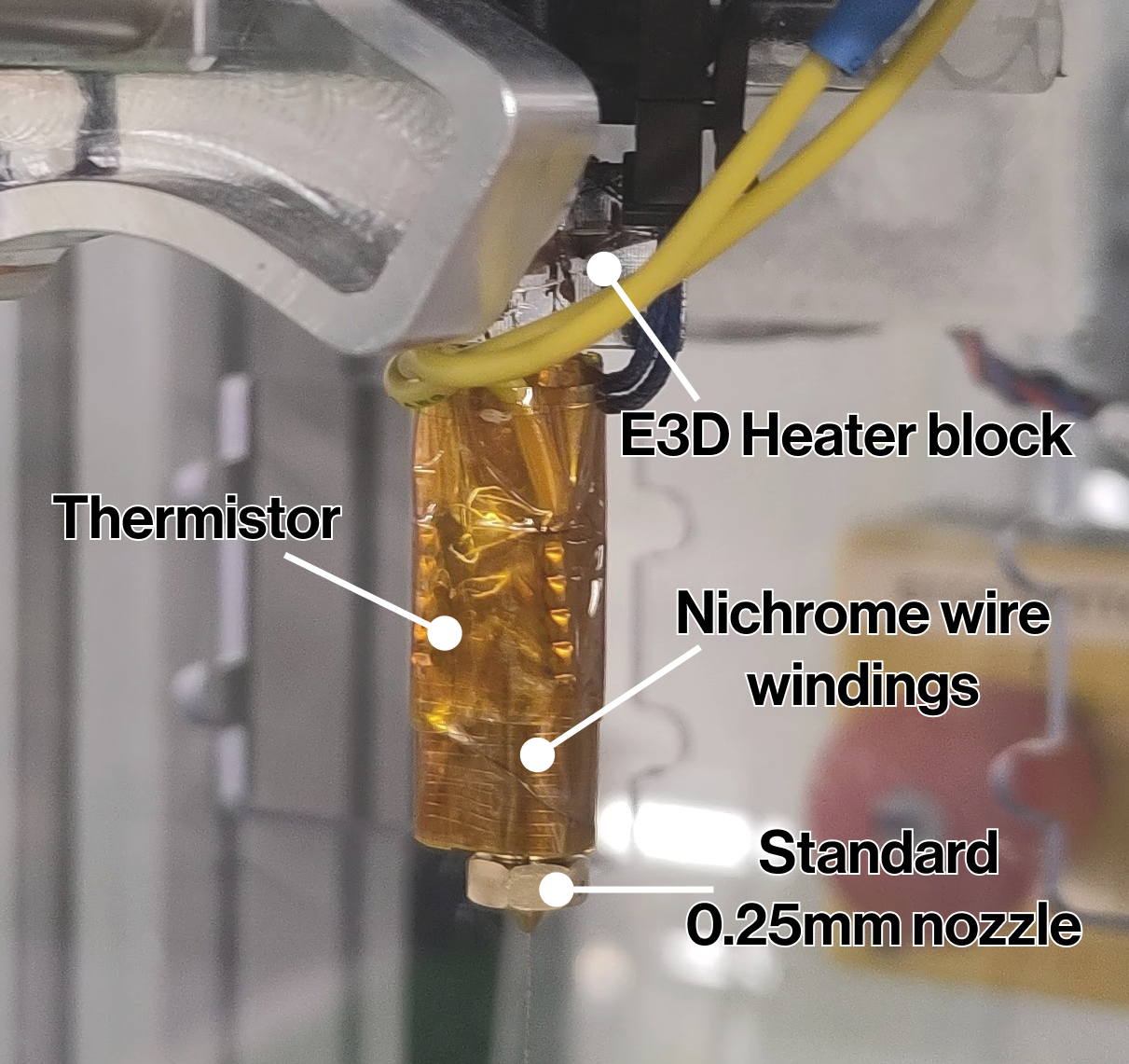}
    \caption{Nozzle extender with nichrome heating element and standard nozzles.}
    \label{fig:nozzle}
\end{figure}

Printing within highly convex geometries requires a thin, elongated nozzle to maximise clearance and avoid collisions. Standard nozzles are typically short to maintain thermal coupling with the heat block. Although extended nozzles are commercially available (e.g., from nonplanar.xyz), the range of available diameters is limited. A custom nozzle extender was therefore designed to interface between the heat block and standard nozzles, enabling the use of widely commercially available smaller diameter nozzles. However, increasing the distance between the heat source and the nozzle tip introduces a significant thermal gradient, leading to insufficient filament melting at the extrusion point and resulting in issues such as clogging and filament stripping. To mitigate this, a heating system based on nichrome wire was implemented. The wire is wrapped around the extender and insulated with Kapton tape (Fig.~\ref{fig:nozzle}). A thermistor is embedded within the nichrome winding, in direct contact with the extender, to enable localised temperature control. This configuration ensures an adequate nozzle-tip temperature for reliable conformal material deposition using a 0.25~mm nozzle, including for the challenging Electrifi filament.

\subsection{Fabrication Data Generation}

The fabrication process begins with a fully defined multi-material 3D model, as required in conventional multi-material MEX workflows. The model should be segmented into distinct solid bodies according to their target materials. In the context of conformal electronics, these can be broadly categorised as substrate geometries and trace (circuit) geometries. Substrate geometries are typically composed of dielectric materials and serve both structural and electromagnetic functions, supporting the circuit and influencing field distribution. Trace geometries, in contrast, are conductive and form the functional electrical pathways, interfacing with other circuit elements and components. While conductive traces dominate performance in DC applications, substrate design becomes increasingly critical at higher frequencies. A manufacturing strategy must then be defined, including planar versus conformal operations and their execution order. Additionally, print orientation and bed placement must be carefully considered. These factors are particularly critical in multi-axis printing, as poor choices can lead to inefficient material usage, degraded surface quality, or cause collisions between the tool-head, printed part or gantry.

\subsubsection{Toolpath Generation}

Geometries intended for conformal deposition require the intermediate step of explicit toolpath generation. The Open5x G-code generator requires both a target surface and a toolpath. However, generating this toolpath remains a largely manual process. This reflects a core difficulty in developing general‑purpose multi‑axis slicers. The additional degrees of freedom transform toolpath planning into an under-constrained optimisation problem. Alternative approaches exist. For instance, the S3 Slicer~\cite{zhang_s3-slicer_2022} demonstrates automated multi-axis slicing capabilities. However, limitations in surface quality and lack of multi-material support make it unsuitable to-date. Similarly, tools such as Fractal Cortex\footnote{https://github.com/fractalrobotics/Fractal-Cortex} may offer partial solutions, but the tightly integrated software and lack of native multi-material workflows introduce some practical challenges. These include reliance on workaround strategies like dummy geometries and reduced control over fine features such as conductive traces. In the context of small conformal electronics, where feature sizes are often on the order of the nozzle diameter (0.1–0.4 mm), robust and fully automated toolpath generation would require a dedicated solution. MM-Open5x addresses this challenge by leveraging the flexibility of Rhino and Grasshopper. This allows users to directly generate and modify toolpaths alongside the G-code generation. While this approach somewhat sacrifices integration and ease of use, it provides a high degree of control, which is essential for research and prototyping. The repository associated with this paper includes Grasshopper scripts (e.g., Fig.~3) used to parametrically generate toolpaths for the case studies presented in Section IV.

\subsubsection{G-code generation}
The final stage consists of generating the manufacturing data (G-code). Planar geometries can be processed using conventional MEX slicers. However, when integrating with the Open5x workflow, it is critical to ensure consistent spatial referencing between tools. In particular, the planar geometry must be positioned at the exact same X–Y location as its counterpart in the Rhino environment used for conformal G-code generation. This is necessary because typical slicers like PrusaSlicer redefine object coordinates based on bounding-box centring, whereas the Open5x generator relies on the global coordinate system of the Rhino workspace for kinematic transformations. Even small misalignments (on the order of half a layer height) can significantly degrade print quality in multi-axis operations.

Once the planar G-code is generated, the conformal toolpaths and target surfaces are used within the Open5x generator to produce the corresponding multi-axis G-code. This is typically exported as a separate file and subsequently merged manually with the planar G-code. Care must be taken to correctly configure the transition between these segments. In particular, the starting G-code for the conformal section must safely position the tool above the part and approach the first toolpath point without collision. Tool-change commands must be properly defined, and for conductive materials such as Electrifi, cold extrusion settings may need to be enabled.

\subsection{In-Process Behaviour and Post-Processing}

The experimental, low‑cost nature of the hardware introduces a number of practical in‑process behaviours that are important to account for when applying the MM‑Open5x workflow. Mechanical tolerances associated with the elongated nozzle extender were found to be particularly sensitive. Inadequate thread precision can lead to leakage of molten filament at the heat block interface, altering extrusion behaviour and degrading surface quality through unintended material deposition. While temporary mitigation (e.g., PTFE tape) can reduce leakage, more reliable operation may be achieved by higher‑tolerance fabrication of the extender.


Bed adhesion is another critical factor in multi‑axis printing. The MM‑Open5x platform does not employ a heated bed or same‑material adhesion surface, with the build plate fabricated from a high‑temperature resin that provides adequate intrinsic adhesion for many geometries. For parts with small contact areas or subjected to significant lateral forces during rotation, additional adhesion was required, typically using PVA‑based adhesives. In conformal printing, nozzle‑induced forces can be sufficient to detach inadequately bonded parts, even in the absence of collisions.

When executing multi‑axis G‑code for previously untested geometries, conservative motion parameters were essential. Initial trials were performed at reduced speeds (e.g., 2~mm/s) and without extrusion. Potential risk regions, typically corresponding to large rotational excursions, were identified and evaluated prior to printing. Cautious verification remains an important component of reliable operation, and some manual intervention during printing is to be expected.



Post‑processing and downstream assembly were intentionally treated as application‑specific and are not standardised within the MM‑Open5x workflow. 
Electroplating of Electrifi traces and conductive component bonding have been explored in prior work~\cite{hong_5-axis_2023,revenga_riesco_multi-axis_2025}, with detailed protocols reported elsewhere~\cite{kim_one-step_2019}. In general, bonding and encapsulation strategies depend strongly on target application requirements and material combinations, and therefore remain beyond the scope of automation within the present workflow.



\begin{figure}[t]
    \centering
    \includegraphics[width=\linewidth]{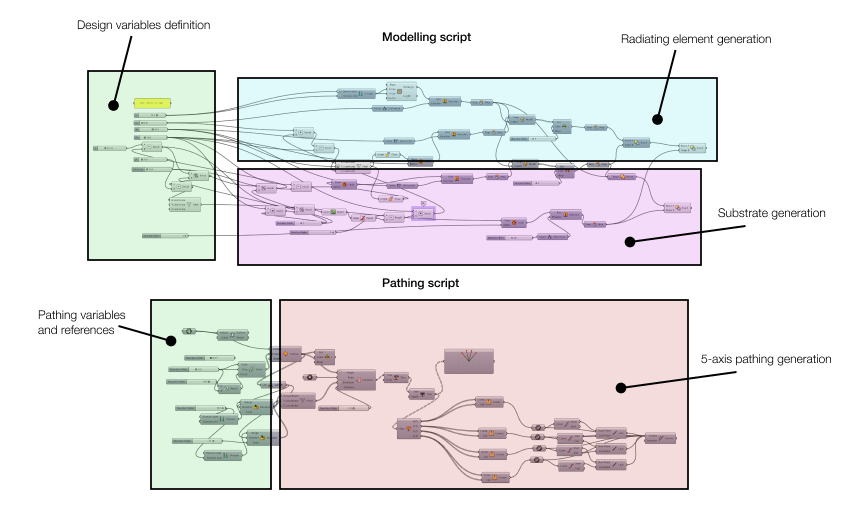}
    \caption{Sample Grasshopper script including parametric 3D model definiton and conformal pathing.}
    \label{fig:patch}
\end{figure}

\section{Example case studies}

This section presents three open‑source case studies that can be found in the repository. These case studies were selected to span RF, wideband, and low‑frequency mixed‑signal electronics, illustrating the applicability and limitations of MM‑Open5x across distinct operating regimes.

\subsection{Conformal Patch Antenna}
\label{patch}
A conformal S-band antenna (Fig.~\ref{fig:patch}, right), originally proposed in \cite{muntoni_curved_2020}. As opposed to the original work, the radiation element is printed conformally on the substrate. This printing orientation showed a closer match to HFSS simulations than when printed planarly~\cite{riesco_5-axis_2025}. The supporting repository contains Ansys HFSS simulation files, including a parametric definition of the antenna within HFSS, as well as a parametric definition of the antenna in Grasshopper. There are also custom Grasshopper pathing scripts, 3D models and the latest G-code.

\begin{figure}[t]
    \centering
    \includegraphics[width=\linewidth]{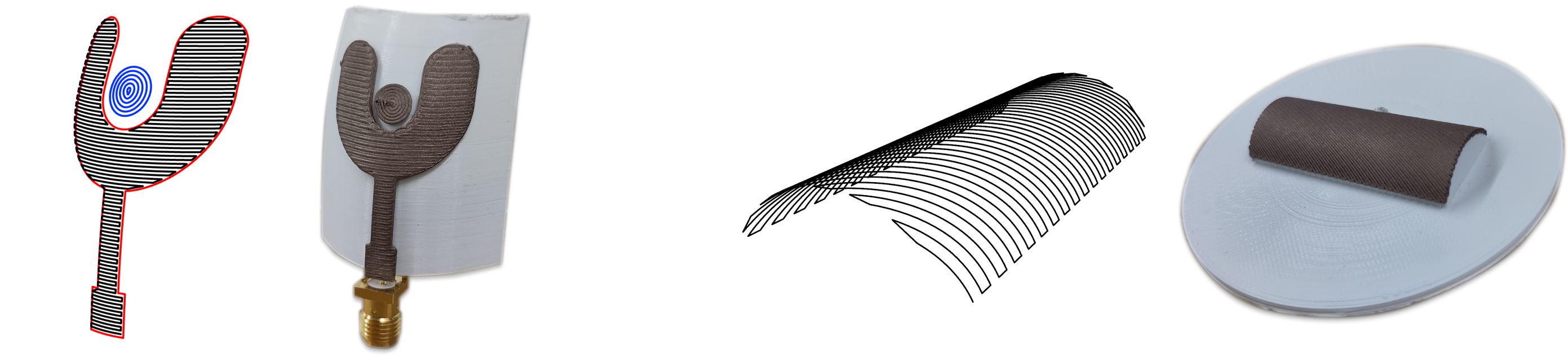}
    \caption{Conformal path of the UWB antenna and fabricated conformal UWB antenna (left); Conformal path of S-band antenna and fabricated conformal S-band antenna (right)~\cite{riesco_5-axis_2025}.}
    \label{fig:patch}
\end{figure}

\subsection{Double-Curvature UWB Antenna}
A double curved conformal Ultra-Wideband Antenna (Fig.~\ref{fig:patch}, left), inspired from \cite{balderas_low-profile_2019}. In the original work it was projected on a single curvature surface. Measurements and tests showed good performance over the printed samples, although high variability and differences with simulations \cite{riesco_5-axis_2025}. Simulation, design, and manufacturing artefacts are provided per Sec.~\ref{patch}.


\subsection{Conformal DC-DC switching regulator}
\begin{figure}[t]
    \centering
    \includegraphics[width=0.7\linewidth]{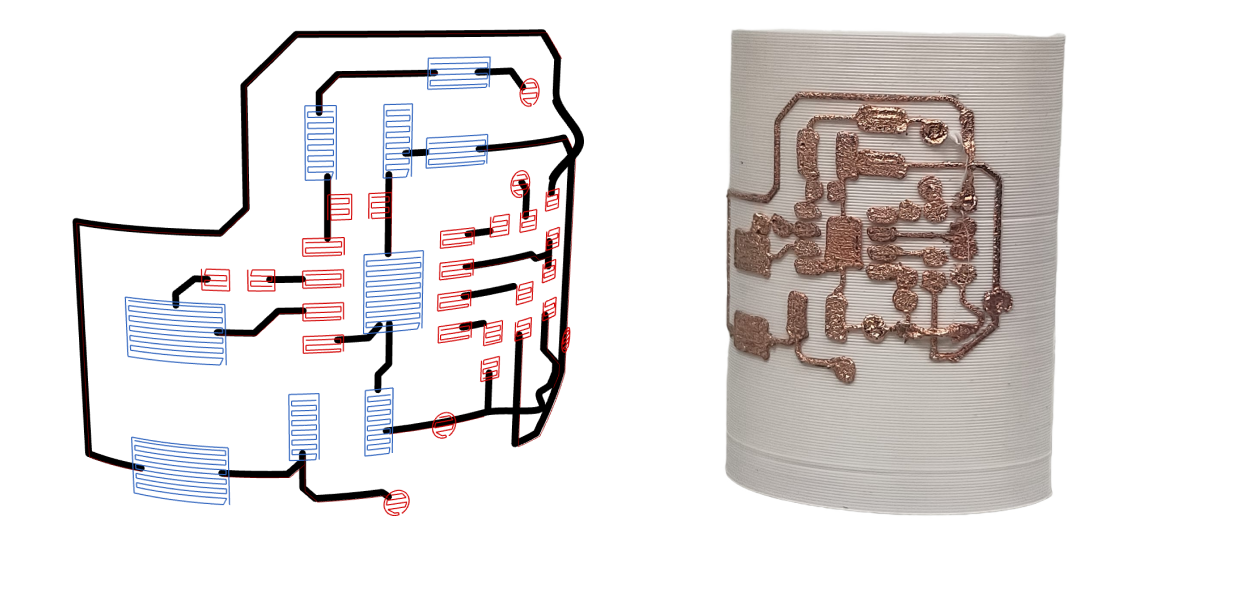}
    \caption{Conformal paths of DC-DC switching regulator (left), Fabricated and plated conformal DC-DC switching regulator (right)~\cite{revenga_riesco_multi-axis_2025}.}
    \label{fig:dcdc}
\end{figure}

A conformal DC-DC regulator (Fig.~\ref{fig:dcdc}), presented in \cite{revenga_riesco_multi-axis_2025}. This case study demonstrates high‑density conformal SMD routing, including bridges and reduced trace clearances, on a low‑cost platform. The repository includes KiCad schematics and PCB files, BOM, Substrate 3D models and G-code for traces, pads and bridges. 

\section{Future Directions and Opportunities}


Despite recent progress, several system‑level challenges remain that offer opportunities for community development. 

\textbf{Component assembly} remains a primary bottleneck for conformal electronics; existing pick‑and‑place systems assume planar substrates and are poorly suited to curved or multi‑orientation geometries, motivating new approaches to conformal‑aware placement, fixturing, and vision. 

\textbf{Low‑temperature bonding} methods enable component integration on polymeric substrates but introduce trade‑offs in reliability, reworkability, and stability. Developing repairable and reversible bonding techniques would significantly benefit prototyping, sustainability, and deployment in long‑lived or resource‑constrained systems.

While the core fabrication workflow is broadly applicable, design constraints can vary significantly across application domains such as wearables, IoT, robotics or space systems. This suggests a need for application‑specific design rules, validation strategies, and reliability metrics that may leverage a shared fabrication foundation. 

\textbf{Sustainability and reuse} present additional opportunities. Choices around materials, bonding and assembly strongly influence repairability, component recovery, and waste, particularly in low‑volume, remote and/or resource-constrained manufacturing contexts.

Finally, open artefacts and shared benchmarks are critical for reproducibility. In addition to releasing simulation and design files, toolpaths, and G‑code, there is value in community‑developed benchmark structures that intentionally stress known failure modes. In practice, \textbf{material variability} across nominally identical batches was found to significantly affect printability and mechanical behaviour, often dominating outcomes despite unchanged parameters. Lightweight acceptance tests and shared reporting of material behaviour represent practical avenues for mitigating these effects and improving inter‑stakeholder reproducibility.


\printbibliography
\balance
\end{document}